\documentclass[letter,aps,prl,showpacs,twocolumn,superscriptaddress]{revtex4-2}

\usepackage[utf8]{inputenc}  
\usepackage[T1]{fontenc}     
\usepackage[british]{babel}  
\usepackage{lmodern}  
\usepackage[scaled=1.03]{inconsolata} 
\usepackage[usenames,dvipsnames]{color} 
\usepackage[colorlinks,citecolor=blue,linkcolor=magenta,urlcolor=blue]{hyperref}  
\usepackage{graphicx} 
\usepackage{tikz}
\usepackage[babel]{microtype}  
\usepackage{physics}
\usepackage{commath}
\usepackage{amsmath,amssymb,amsthm,bm,mathtools,amsfonts,mathrsfs,bbm,dsfont} 
\usepackage{xspace}  
\usepackage{multirow}
\usepackage{enumitem}
\usepackage{verbatim}
\usepackage{float}
\usepackage{kbordermatrix}
\usepackage{bbold}
\usepackage{tcolorbox}
\usepackage{cleveref}

\usepackage{dblfloatfix}
\usepackage{blindtext}
\usepackage{xcolor}

\renewcommand{\kbldelim}{(}
\renewcommand{\kbrdelim}{)}

\newcommand{\id}{\ensuremath{\mathds{1}}}
\renewcommand{\vec}[1]{\boldsymbol{#1}}
\newcommand{\floor}[1]{\left \lfloor #1 \right \rfloor}

\let\norm\undefined
\DeclarePairedDelimiter\norm{\lVert}{\rVert}

\newcommand{\R}{\mathbb{R}}

\newcommand{\cD}{\mathcal{D}}
\newcommand{\cZ}{\mathcal{Z}}

\newcommand{\cF}{\mathcal{F}}

\newcommand{\defeq}{\vcentcolon=}

\newtheoremstyle{mystyle}
  {6pt}
  {6pt}
  {\normalfont}
  {0pt}
  {\bf}
  {.}
  { }
  {}

\theoremstyle{mystyle}
\newtheorem{theorem}{Theorem}
\newtheorem{conjecture}{Conjecture}
\newtheorem{proposition}{Proposition}
\newtheorem{lemma}{Lemma}

\newtheorem{corollary}{Corollary}

\newtheorem{remark}{Remark}

\newenvironment{reusethm}[1]{%
  \renewcommand\thetheorem{\ref{#1}}\theorem}{\endtheorem\addtocounter{theorem}{-1}}

\colorlet{myPurple}{blue!40!red}
\colorlet{myCyan}{cyan!50!gray}

\definecolor{quantumviolet}{HTML}{53257F} 
\definecolor{quantumgray}{HTML}{555555} 
\definecolor{mygray}{gray}{0.95} 

\newtcolorbox[auto counter,number within=section]{boxfigure}[2][]{%
colback=mygray,colframe=myPurple,fonttitle=\bfseries,width=\columnwidth,float*=ht,lower separated=false, halign=justify,title=Box~\thetcbcounter: #2,#1}

\begin{document}
\nonfrenchspacing

\title{On compatibility of binary qubit measurements}

\author{Dmitry Grinko}
\affiliation{QuSoft, University of Amsterdam, Amsterdam, The Netherlands}
\affiliation{Department of Mathematics, University of Geneva, 1211 Geneva, Switzerland}
\author{Roope Uola}
\affiliation{Department of Applied Physics, University of Geneva, 1211 Geneva, Switzerland}

\date{\today}

\begin{abstract}
Deciding which sets of quantum measurements allow a simultaneous readout is a central problem in quantum measurement theory. The problem is relevant not only from the foundational perspective, but also has direct applications in quantum correlation problems that are fueled by the presence of incompatible measurements. Although central, only few analytical criteria exist for deciding the incompatibility of general sets of measurements. In this work, we approach the problem through functions defined on the Boolean hypercube and their Fourier transformations. We show that this reformulation of the problem leads to a complete geometric characterisation of joint measurability of any finite set of unbiased binary qubit measurements, and gives a necessary condition for the biased case. We discuss our results in the realm of quantum steering, where they translate into a family of steering inequalities. When certain unbiasedness conditions are fulfilled, these criteria are tight, hence fully characterizing the steering problem, when the trusted party holds a qubit and the untrusted party performs any finite number of binary measurements. We further discuss how our results point towards a second-order cone programming approach to measurement incompatibility, and compare this to the predominantly used semi-definite-programming-based techniques. We use our approach to falsify an existing conjecture on measurement incompatibility of special sets of measurements.
\end{abstract}

\maketitle

\section{Introduction} 

Quantum measurement incompatibility forms a cornerstone of quantum information theory. In addition to its importance in describing foundational aspects of quantum theory in, e.g., the Heisenberg microscope \cite{buschrmp2014} and fundamental interferometric bounds \cite{Kiukas2022}, incompatibility has recently found various applications in quantum information theory \cite{JMReviewGuhne}. To mention a few, incompatibility is known to fully characterize quantum steering \cite{marco2014JM,uola2014JM,uola2015JM,kiukas2017CV}, instances of Bell non-locality \cite{wolf09,quintino15,bene18,plavala2024IncmultiBell}, as well as preparation contextuality \cite{tavakoli2020JM,selby21,Plavala2022GPTJMPNC}, and it works as a prerequisite for the quantum advantage in various prepare-and-measure scenarios \cite{skrzypczyk19,carmeli19a,oszmaniec19,uola19b,uola19c,guerini19,buscemi20,deGois2021,Saha2023,masini2024jointmeasurability}.

The fundamental interest on and the wide applicability of incompatibility have motivated a long line of research on characterizing and quantifying this non-classical phenomenon. This includes the development of semi-definite programming techniques \cite{wolf09,Heinosaari2015b,Skrzypczyk15,uola2015JM,cavalcanti17,designolle19a}, quantification techniques motivated by resource theoretic measures \cite{pusey15,haapasalo15a,uola2015JM,designolle19b}, (semi-)device independent witnesses \cite{chen16b,cavalcanti16,quintino19,chen21}, geometric methods \cite{busch1986unsharp,brougham2007,stano2008coexistence,busch2010coexistence,yu2010joint,pal2011approximate,liang2011specker,yu2013quantum,uola2016adaptive,andrejic2020joint,jae2019tertiary}, Banach space techniques \cite{bluhmJointMeasurabilityQuantum2018,bluhmCompatibilityQuantumMeasurements2020,faedi21}, and formulations of the concept in generalized probabilistic theories \cite{Plavala2017,Filippov2018,Bluhm2022}, as well as experimental quantification of the effect \cite{Hammad2020}. We note that all the mentioned works are in the finite-dimensional and discrete outcome setting, and refer to the work of Busch et al. for a collection of results in the continuous realm \cite{busch2016quantum}.

In this manuscript, we introduce an analytic method for characterizing incompatibility of yes-no questions, i.e. binary measurements, based on Fourier transformations of functions defined on the Boolean hypercube. This amounts to representing the existence problem of joint measurements through Fourier transformations of their Bloch representation, and constructing a solution for this dual problem. By utilizing the theory of Fourier analysis \cite{de2008brief,o2014analysis}, we derive a collection of necessary criteria for compatibility of any finite number of binary qubit measurements. We show that when the involved measurements are unbiased, i.e. projective measurements subject to white noise, the criteria give a full analytical characterization of incompatibility. This generalizes various known special instances of the geometric approach to compatibility and as such, we give a brief overview of the existing results on this research line. As direct applications, we give a characterization of the related quantum steering problems, discuss the possibility of using second-order cone programming for solving the problem of joint measurability, and falsify a conjecture on measurement incompatibility presented in \cite{andrejic2020joint}.

\section{Joint measurability}

We model quantum measurements as positive operator valued measures (POVMs). These are collections of positive semi-definite Hermitian operators $\{J(\nu)\}_{\nu \in \Omega}$ acting on a finite-dimensional Hilbert space summing to the identity operator, i.e. $\sum_{\nu \in \Omega} J(\nu)=\id$, where $\Omega$ is the set of outcome labels. We denote a measurement assemblage by $\mathcal M=\{J_{i}(\mu_i)\}_{\mu_i,i}$, where $i \in \{1,\dotsc,N\}$ labels the choice of a measurement, i.e. $\{J_{i}(\mu_i)\}_{\mu_i \in \Omega_i}$ is a POVM for each $i$. For an assemblages of $N$ POVMs measurement compatibility is defined through the concept of joint measurability \cite{busch2016quantum,heinosaari16b,JMReviewGuhne}. An assemblage $\mathcal M$ is called \emph{jointly measurable}, if there exists a single POVM $\{J(\mu)\}_{\mu \in \Omega}$ with $\mu = (\mu_1,\dotsc,\mu_N)$ and $\Omega = \Omega_1 \times \dotsc \times \Omega_N $ such that
\begin{equation}
    J_{i}(\mu_i)=\sum_{\mu_1,\dotsc,\mu_{i-1},\mu_{i+1},\dotsc,\mu_N} J(\mu_1,\dotsc,\mu_N)
\end{equation}
for all $i$ and $\mu_i$. In this case, the POVM $\{J(\mu)\}_{\mu \in \Omega}$ is called a joint measurement of the measurement assemblage $\mathcal M=\{J_{i}(\mu_i)\}_{\mu_i,i}$. If such POVM does not exits, the measurement assemblage is called incompatible.

As an illustrative example, consider two noisy spin measurements given by the qubit POVMs $J^\eta_{x}(\pm)=\frac{1}{2}(\id \pm\eta\sigma_x)$ and $J^\eta_{z}(\pm)=\frac{1}{2}(\id \pm\eta\sigma_z)$, where $\eta$ quantifies the amount of noise. An intuitive way of building a joint measurement is to flip a coin between measuring a sharp spin observable in the directions $x+z$ and $x-z$, and interpreting the outcomes as joint outcomes of the noisy measurements. For example, the plus outcome of $x+z$ is interpreted as both noisy measurements having the outcome plus. This procedure is described by the POVM
\begin{align}
    J(\mu_1,\mu_2)=\frac{1}{4}\Bigg[\openone+\frac{1}{\sqrt 2}(\mu_1\sigma_x+\mu_2\sigma_z)\Bigg],
\end{align}
where $\mu_1,\mu_2\in\{-1,1\}$. Clearly the marginals of this POVM are the noisy spin measurements with $\eta=1/\sqrt 2$. Indeed, one can prove that this bound is optimal in that for any $\eta\leq1/\sqrt 2$ one has joint measurability and for any $\eta>1/\sqrt 2$ the noisy measurements $J^\eta_{x}(\pm), J^\eta_{z}(\pm)$ are incompatible \cite{busch1986unsharp}. We note that for $\eta\neq0$ the noisy measurements do not commute with each other. In other words, joint measurability and commutativity are not equivalent concepts for general measurements. However, commutativity does imply joint measurability by using the product of the involved POVMs as the joint measurement.

\section{A brief review on known geometric results in the qubit case} 

Take two binary qubit POVMs $J_1$ and $J_2$ given by the elements $J_i(\pm)=\frac{1}{2}\big((1\pm z_i)\pm\bm{z}_i\cdot\bm{\sigma}\big)$, where the scalar $z_i$ is called the bias and $\bm{z}_i$ is the Bloch vector. One way to interpret the bias term is to note that an unbiased measurement on a maximally mixed state results in a uniform outcome probability distribution, while this is not the case for a biased measurement. Busch proved \cite{busch1986unsharp} that in the unbiased case, i.e. when $z_i=0$ for every $i$, joint measurability of such pairs is equivalent to
\begin{equation}\label{eq:Busch}
    \norm{\bm{z}_1 - \bm{z}_2} + \norm{\bm{z}_1 + \bm{z}_2} \leq 2,
\end{equation}
where $\norm{\cdot}$ is the usual Euclidean norm on $\R^{3}$.

The proof of this fact is based on parametrizing all possible joint measurements in the Bloch picture. The existence of a joint measurement is then shown to be equivalent to an empty intersection property of balls drawn around the Bloch vectors $0,\bm{z}_1,\bm{z}_2$ and $\bm{z}_1+\bm{z}_2$. The proof of our result presents a generalization of this argument, where the joint measurements are parametrized through Fourier transformations of functions defined on the Boolean hypercube and the existence of the joint measurement is related to the properties of Fourier coefficients. These coefficients do not appear in the above criterion, as in the case of two measurements the relevant coefficients can be eliminated, as we discuss after Theorem \ref{thm:necessary_condition}.

These results were extended to the case of three binary qubit measurements by the authors of \cite{pal2011approximate}, who provided the following necessary condition for joint measurability in the biased case, i.e. when the bias term $z_i$ is non-zero,
\begin{align}\label{Eq:FermatTorricelli}
\sum_{i=0}^3\norm{\bm{x}_i - \bm{x}_{FT}} \leq 4.
\end{align}
Here, $\bm{x}_0=\bm{z}_1+\bm{z}_2+\bm{z}_3$, $\bm{x}_i=2\bm{z}_i-\bm{x}_0$ for $i=1,2,3$, and we have introduced $\bm{z}_3$ to refer to the Bloch vector related to the third measurement. Here the point $\bm{x}_{FT}$ is the Fermat-Torricelli point of the involved vectors $\bm{x}_i$. This is the point that minimises the sum of distances in the expression in Eq.~\ref{Eq:FermatTorricelli}. The above condition was proven to be sufficient for the unbiased case by Yu and Oh in Ref.~\cite{yu2013quantum}. We note that this joint measurability criterion includes the special case of three orthogonal measurements originally discussed in \cite{busch1986unsharp,brougham2007} and the case of three qubit measurements whose Bloch vectors are equally spaced on the Bloch sphere \cite{liang2011specker}. As another special case, when one has only two measurements, i.e. $\bm{z}_3=0$, this gives the Busch criterion of Eq.~(\ref{eq:Busch}). In our analysis, where we relate the existence of a joint measurement to properties of Fourier coefficients of functions defined on the Boolean hypercube, the case of three measurements gives one Fourier coefficient that can not be eliminated. This coefficient is exactly the Fermat-Torricelli point, as we show below Theorem \ref{thm:necessary_condition}.

On top of the general characterizations for two and three binary measurements, tight geometric criteria for symmetric sets of unbiased binary qubit measurements were derived in \cite{uola2016adaptive}, and various special sufficient, as well as necessary conditions were obtained in \cite{andrejic2020joint}, with the main focus being in measurements with Bloch vectors laying in the same plane. Criteria in a similar spirit also exist for the case of two trinary qubit measurements \cite{jae2019tertiary}, and for all noisy projective qubit measurements \cite{uola2014JM}.

Finally, for biased measurements fewer characterizations are known, while many of the known unbiased criteria are necessary conditions for compatibility also in the biased case. However, for the case of two biased binary qubit measurements, a geometric characterisation of joint measurability was proven independently in Refs.~\cite{stano2008coexistence,busch2010coexistence,yu2010joint}. In this case, the criterion reads
\begin{align}\label{eq:pairbias}
    \big(1-F_1^2-F_2^2\big)\big(1-\frac{\gamma_1}{F_1}-\frac{\gamma_2}{F_2}\big)\leq(\bm{z}_1\cdot\bm{z}_2-z_1 z_2),
\end{align}
where $F_i=\frac{1}{2}\Big[\sqrt{(1+z_i)^2-\|\bm{z}_i\|}+\sqrt{(1-z_i)^2-\|\bm{z}_i\|}\Big]$.

\section{Main Results}

We extend the Bloch-vector-based approach discussed in the previous section to the case of any finite set of binary qubit POVMs. The resulting criteria are necessary for both the biased and the unbiased case. Thus, violations of these criteria witness measurement incompatibility. Moreover, in the case of unbiased measurements, the criteria are tight. As the proof constitutes of putting together a collection of smaller observations, we present the main result with some examples in the main text, and present the detailed steps of the proofs in the Appendix. Our approach is inspired by \cite{yu2013quantum}.

We note that any joint measurement $J$ of $N$ binary POVMs can be seen as a function from a Boolean cube $\Omega^N=\{-1,+1\}^N$ to positive semi-definite matrices. In the qubit case, we further have
\begin{equation}
    J(\mu) = \hat{z}(\mu) \id + \hat{\bm{z}}(\mu) \cdot \vec{\sigma}
\end{equation}
for some functions $\hat{z} : \Omega^N \rightarrow \mathbbm{R}$ and $\hat{\bm{z}} : \Omega^N \rightarrow \mathbbm{R}^3$. Taking the Fourier transformation of the first function \cite{de2008brief,o2014analysis}, we have
\begin{equation}
    \hat{z}(\mu) = \frac{1}{2^N} \sum_{S \subseteq [N]} z(S) \chi_S(\mu),
\end{equation}
where $\chi_S(\mu) \defeq \prod_{i \in S} \mu_i$ for every nonempty $S \subseteq [N]$ and $\chi_\varnothing(\mu) \defeq 1$. The similar expression is true for vector-valued function $\hat{\bm{z}}(\mu)$. We show in the Appendix (Prop.~\ref{prop:fourier}), that degree one Fourier coefficients of such functions coincide with the Bloch vectors of the corresponding marginal POVMs $J_i$. In other words, $\bm{z}(\{i\}) = \bm{z}_i$, where $\bm{z}_i$ are Bloch vectors of POVMs $J_i$. Enforcing functions $ \hat{z}(\mu), \hat{\bm{z}}(\mu)$ to represent a POVM, i.e. to have positive semi-definite values, translates into an inequality involving the Fourier coefficients and the corresponding biases, cf. Eq.~(\ref{thm:eq:positivity}). Putting all this together one gets the following necessary condition for joint measurability, cf. Appendix for its detailed derivation. We first express the technical result and explain it with concrete examples straight after.

\begin{theorem} \label{thm:necessary_condition}
If $N$ binary qubit observables, specified by the Bloch vectors $\bm{z}_1,\dotsc,\bm{z}_N$, are jointly measurable, then
\begin{equation} \label{eq:necessary_condition}
    2^{N-1} \geq \min_{\cF_{N}} \sum_{\mathbf{\mu} \in \cD^{N}} \norm[\Bigg]{ \sum_{\substack{S \subseteq [N] \\ \text{odd }\abs{S}}} \bm{z}(S) \chi_S(\mu) },
\end{equation}
where the configuration set $\cD^{N}$ is defined as 
\begin{equation}
        \cD^{N} \defeq 
\left\{ \mu \in \Omega^N \; \middle \vert \; \mu_1 = +1 \right\}.
\end{equation}
The optimisation is done over the set $\cF_{N}$ of $\mathbbm{R}^3$-vectors defined as 
\begin{equation}
    \cF_{N} \defeq \left\{ \bm{z}(S) \in \mathbbm{R}^3 \; \middle \vert \; S \subseteq [N] \text{, } \abs{S} \text{ is odd, } \abs{S} \geqslant 3 \right\}.
\end{equation}
The vectors $\bm{z}(S)$ with $\abs{S} = 1$ correspond to the Bloch vectors of the given $N$ qubit POVMs. We use notation $\bm{z}_i$ as a shorthand for $\bm{z}(\{i\})$ and, similarly, $\bm{z}_{i \dotsc k}$ as a shorthand for $\bm{z}(\{i,\dotsc,k\})$
\end{theorem}

The condition in Eq.~(\ref{eq:necessary_condition}) and the optimization set $\cF_{N}$ are best explained by examples. First, when $N=2$, we get $\cD^{2}=\{(+,+),(+,-)\}$ and there is no optimisation as the set $\cF_2 = \varnothing$. Hence, Eq.~(\ref{eq:necessary_condition}) becomes
\begin{align}
    2 &\geq \sum_{\mathbf{\mu} \in \cD^{2}} \norm[\Bigg]{ \sum_{\substack{S \subseteq [2] \\ \text{odd }\abs{S}}} \bm{z}(S) \chi_S(\mu) } = \sum_{\mathbf{\mu} \in \cD^{2}} \norm[\Bigg]{ \sum_{i=1}^{2} \bm{z}_i \mu_i } =  \nonumber \\
    &= \norm{\bm{z}_1 - \bm{z}_2} + \norm{\bm{z}_1 + \bm{z}_2}, 
\end{align}
i.e. we recover the Busch criterion in Eq.~(\ref{eq:Busch}).

In the case $N=3$, the set $\cF_3$ has one element, i.e. the Fermat-Torricelli point of the triplet of the involved measurements, and one recovers Eq.~(\ref{Eq:FermatTorricelli}):
\begin{align}
    2^2 &\geq \sum_{\mathbf{\mu} \in \cD^{3}} \norm[\Bigg]{ \sum_{\substack{S \subseteq [3] \\ \text{odd }\abs{S}}} \bm{z}(S) \chi_S(\mu) } = \\
    &=\norm{\bm{z}_1 + \bm{z}_2 + \bm{z}_3 + \bm{z}_{123} } +\norm{\bm{z}_1 - \bm{z}_2 + \bm{z}_3 - \bm{z}_{123} } + \nonumber \\ 
    &+\norm{\bm{z}_1 + \bm{z}_2 - \bm{z}_3 - \bm{z}_{123} } +\norm{\bm{z}_1 - \bm{z}_2 - \bm{z}_3 + \bm{z}_{123} }, \nonumber 
\end{align}
where $\bm{z}_1,\bm{z}_2,\bm{z}_3$ are given and $\bm{z}_{123}$ is the Fermat--Torricelli point.

For higher numbers of measurements, the set $\cF_N$ includes $2^{N-1}-N$ points, which will be manifest in the corresponding joint measurability criterion. For example, in the case $N=4$, the set $\cF_4$ includes four elements (Fourier coefficients). In this case, every triplet of measurements gets its own optimisation variable in $\cF_4$, but these do not need to be the Fermat-Torricelli points of the triplets. Namely, the condition reads as
\begin{align}
    2^3 \geq& \sum_{\mathbf{\mu} \in \cD^{4}} \norm[\Bigg]{ \sum_{\substack{S \subseteq [4] \\ \text{odd }\abs{S}}} \bm{z}(S) \chi_S(\mu) } \\
    =&\norm{\bm{z}_1 + \bm{z}_2 + \bm{z}_3 + \bm{z}_4 + \bm{z}_{234} + \bm{z}_{134} + \bm{z}_{124} + \bm{z}_{123}} \nonumber \\ 
    +&\norm{\bm{z}_1 - \bm{z}_2 + \bm{z}_3 + \bm{z}_4 - \bm{z}_{234} + \bm{z}_{134} - \bm{z}_{124} - \bm{z}_{123}} \nonumber \\
    +&\norm{\bm{z}_1 + \bm{z}_2 - \bm{z}_3 + \bm{z}_4 - \bm{z}_{234} - \bm{z}_{134} + \bm{z}_{124} - \bm{z}_{123}} \nonumber \\
    +&\norm{\bm{z}_1 + \bm{z}_2 + \bm{z}_3 - \bm{z}_4 - \bm{z}_{234} - \bm{z}_{134} - \bm{z}_{124} + \bm{z}_{123}} \nonumber \\
    +&\norm{\bm{z}_1 + \bm{z}_2 - \bm{z}_3 - \bm{z}_4 + \bm{z}_{234} + \bm{z}_{134} - \bm{z}_{124} - \bm{z}_{123}} \nonumber \\
    +&\norm{\bm{z}_1 - \bm{z}_2 + \bm{z}_3 - \bm{z}_4 + \bm{z}_{234} - \bm{z}_{134} + \bm{z}_{124} - \bm{z}_{123}} \nonumber \\
    +&\norm{\bm{z}_1 - \bm{z}_2 - \bm{z}_3 + \bm{z}_4 + \bm{z}_{234} - \bm{z}_{134} - \bm{z}_{124} + \bm{z}_{123}} \nonumber \\
    +&\norm{\bm{z}_1 - \bm{z}_2 - \bm{z}_3 - \bm{z}_4 - \bm{z}_{234} + \bm{z}_{134} + \bm{z}_{124} + \bm{z}_{123}}, \nonumber
\end{align}
where $\bm{z}_1,\bm{z}_2,\bm{z}_3,\bm{z}_4$ are the Bloch vectors of the given measurements, and one gets the additional Fermat-Torricelli-type points $\bm{z}_{234}, \bm{z}_{134},\bm{z}_{124}, \bm{z}_{123}$ as variables.

We note that the necessary condition is valid for both biased and unbiased measurements. In the following, we state that the criterion is tight in the case of unbiased measurements. The proof is based on well-chosen ansatz for the joint POVM $J$, which we give in the Appendix (see Eq.~\ref{thm2:def_vec_scal}).

\begin{theorem} \label{thm:sufficient_condition}
If for $N$ binary unbiased qubit observables, specified by vectors $\bm{z}^\prime_1,\dotsc,\bm{z}^\prime_N$, there exist vectors $\bm{z}^\prime(S) \in \cF_{N}$ for every corresponding odd $\abs{S} > 1$, such that
\begin{equation} \label{eq:sufficient_condition}
    2^{N-1} \geq \sum_{\mathbf{\mu} \in \cD^{N}} \norm[\Bigg]{ \sum_{\substack{S \subseteq [N] \\ \text{odd }\abs{S}}} \bm{z}^\prime(S) \chi_S(\mu) },
\end{equation}
then these $N$ qubit observables are jointly measurable.
\end{theorem}

\section{Quantum steering} 

Joint measurability is closely related to the problem of quantum steering \cite{marco2014JM,uola2014JM,uola2015JM,kiukas2017CV}. Quantum steering asks whether two parties, Alice and Bob, can verify the entanglement of a bipartite state when only one party's measurement apparatuses are trusted, i.e. they have a quantum description. This amounts to falsifying so-called local hidden state models. These are models that aim at explaining locally the changes in Bob's share of a bipartite state when Alice performs measurements, inputs of which are chosen by Bob, on her share of the state, and communicates the outcomes to Bob. Given that Alice performs a POVM $\{A_{i}(\mu_i)\}$ and the shared state is $\varrho_{AB}$, Bob's local subnormalised post-measurement states read $\sigma_{i}(\mu_i):=\text{tr}_A[(A_{i}(\mu_i)\otimes\openone)\varrho_{AB}]$. Such state assemblage $\{\sigma_{i}(\mu_i)\}$ is said to have a local hidden state model if it can be decomposed as \cite{wiseman07,cavalcanti17,uola20review}
\begin{align}
\sigma_{i}(\mu_i)=p(\mu_i|i)\sum_\lambda p(\lambda|\mu_i,i)\sigma(\lambda). 
\end{align}
Here, $p(\mu_i|i)$ is a normalisation factor and $p(\lambda|\mu_i,i)$ are priors for the ensemble $\{\sigma(\lambda)\}$ of subnormalised states that get updated upon the classical communication, i.e. upon Bob knowing the input $i$ and the corresponding output $\mu_i$.

The connection between joint measurability and steering is given by the map $\sigma_{i}(\mu_i)\mapsto\rho_B^{-1/2}\sigma_{i}(\mu_i)\rho_B^{-1/2}=:J_{i}(\mu_i)$, where $\rho_B=\text{tr}_A[\varrho_{AB}]$ and a pseudoinverse is used when necessary. Clearly $\{J_{i}(\mu_i)\}$ forms a set of POVMs, called steering-equivalent POVMs, and it is straight-forward to see \cite{uola2015JM} that $\{\sigma_{i}(\mu_i)\}$ has a local hidden state model if and only if the POVMs $\{J_{i}(\mu_i)\}$ are jointly measurable.

Using the steering-equivalent POVMs for a qubit state assemblage with two outputs and any finite number of inputs, the criteria given in Theorem~\ref{thm:necessary_condition} turns into steering inequalities: if a state assemblage has a local hidden state model, then the steering-equivalent POVMs are jointly measurable and, consequently, the criteria in Eq.~(\ref{eq:necessary_condition}) hold. If, additionally, the steering-equivalent POVMs are unbiased, a full characterisation of the steering problem is achieved for any finite number of inputs due to Theorem \ref{thm:sufficient_condition}. This is the case, for example, when Alice and Bob share a noisy singlet state between two-qubit systems and Alice performs unbiased binary measurements. We summarize this in the following Corollary.

\begin{corollary}
    The criteria given by Eq.~(\ref{eq:necessary_condition}) are steering inequalities for a scenario where the untrusted party performs any finite number of binary measurements and the trusted party holds a qubit. If the steering equivalent POVMs are unbiased, these inequalities fully characterize the steering problem.
\end{corollary}

\section{Numerical method} 

Our characterisation in \Cref{thm:necessary_condition,thm:sufficient_condition} can formulated in the form of \emph{second-order cone program (SOCP)} as follows:
\begin{equation}
\begin{aligned}
\min_{x,t} & \quad e^T t \\
 \text{s.t.} & \quad \norm{A_i x+b_i} = t_i, \quad i = 1, \ldots, M
\end{aligned}
\end{equation}
where $M = 2^{N-1}$, $x \in \mathbbm{R}^{3(M-N)}$, $b_i \in \mathbbm{R}^{3}$, $A_i \in \mathbbm{R}^{3 \times 3(M-N)}$ and $t,e \in \mathbbm{R}^{M}$; $i$ corresponds to $\mu$ and under this equivalence we have
\begin{align}
    e &= (1,\dotsc,1), \quad x = \bigoplus_{\substack{S \subseteq [N] \\ \text{odd }\abs{S}}} (z_1(S),z_2(S),z_3(S)), \\
    b_{\mu} &= \sum_{k=1}^{N} \bm{z}_k \mu_k, \quad A_{\mu}x= \sum_{\substack{S \subseteq [N], \abs{S} > 1 \\ \text{odd }\abs{S} }} \bm{z}(S) \chi_S(\mu).
\end{align}

We have implemented this formulation of joint measurability numerically, and present here an application of it by solving a conjecture put forward in Ref.~\cite{andrejic2020joint}. In Ref.~\cite[Theorem 8]{andrejic2020joint} a sufficient condition for joint measurability of \emph{biased coplanar} binary POVMs is given as
\begin{equation}\label{conj3:eq}
    \norm{\vec{z}_{1}+\vec{z}_{N}} +\sum_{p=1}^{N-1}\norm{\vec{z}_{p}-\vec{z}_{p+1}} \leqslant 2\left(1-\max_{k \in [N]} \left\{\left|b_{k}\right|\right\}\right),
\end{equation}
where $\{ \vec{z}_{k} \}_{k=1}^{N}$ are the Bloch vectors of the given $N$ POVMs and $\{ b_{k} \}_{k=1}^{N}$ are the corresponding biases. \emph{Biased coplanarity} here means that all $\{ \vec{z}_{k} \}_{k=1}^{N}$ belong to the same plane and the POVMs are labelled such that all $b_{k} \geqslant 0$, and that the POVMs themselves are labelled by non-decreasing bias, i.e. $b_{k} \geqslant b_{p}$ for $k \geqslant p$. The authors of Ref.~\cite{andrejic2020joint} present the following conjecture.

\begin{conjecture}\label{conj3}
The sufficient condition~(\ref{conj3:eq}) is necessary in the case of \emph{unbiased coplanar} POVMs, where
\begin{enumerate}[nolistsep]
    \item The Bloch vectors $\vec{z}_{k}$ are arranged by the angles between $\vec{z}_{k}$ and $x$-axis in increasing order;
    \item The Bloch vectors $\vec{z}_{k}$ lie on the lines that make the total angular span less than $\pi$ and they all point towards the upper half of the plane;
    \item None of the Bloch vectors is a convex combination of any other pair of Bloch vectors.
\end{enumerate}
\end{conjecture}

Using our characterisation of joint measurability for unbiased measurements and the numerical technique given by SOCP, we found a counterexample to the Conjecture~\ref{conj3}. Consider $N = 4$ and let
\begin{align}
    \vec{z}_1 &= (0.7, 0.0,0), \nonumber\\
    \vec{z}_2 &= (0.8, 0.4,0), \\
    \vec{z}_3 &= (0.3, 0.4,0), \nonumber\\
    \vec{z}_4 &= (0.1, 0.2,0) \nonumber.
\end{align} 
This set of Bloch vectors satisfies the conditions of the Conjecture~\ref{conj3}. Moreover, it is not difficult to check that:
\begin{equation*}
    2 < \norm{\vec{z}_{1}+\vec{z}_{4}} +\sum_{p=1}^{3}\norm{\vec{z}_{p}-\vec{z}_{p+1}} \approx 2.01977.
\end{equation*}
However, using our notation $\vec{z}(i) \defeq \vec{z}_i$, the condition of the Theorem~\ref{thm:sufficient_condition} is satisfied:
\begin{equation*}
    2^3 > \min_{\cF_{N}} \sum_{\mathbf{\mu} \in \cD^{N}} \norm[\Bigg]{ \sum_{\substack{S \subseteq [N] \\ \text{odd }\abs{S}}} \bm{z}(S) \chi_S(\mu) } \approx 7.95738.
\end{equation*}

On the contrary, we found no counterexamples to other two conjectures presented in Ref.~\cite{andrejic2020joint}  numerically.

\section{Conclusions} 

We have applied Fourier transformations of functions defined on the Boolean hypercube to attack the problem of joint measurability of binary measurements. We have explicitly concentrated on the qubit case, wherein the method provides a full characterisation for joint measurability of an arbitrary (finite) set of unbiased binary measurements. For biased measurements, the resulting criterion is a necessary condition for joint measurability. On the other hand, the unbiased case was only known for pairs of binary and trinary measurements, as well as for general triplets and sets of measurements with symmetries. In contrast, our results fully resolve joint measurability of unbiased binary qubit measurements in any physically reachable scenario, i.e. one with finitely many measurements.

Using the connection between steering and joint measurability, we argued that our results translate to steering criteria in the related quantum steering scenario. Moreover, under certain unbiasedness conditions, these steering criteria are tight, hence, fully characterising the problem.

An open problem is a geometric interpretation of our conditions \Cref{thm:necessary_condition,thm:sufficient_condition}. Namely, our criteria are based on Fermat-Torricelli type points. To our knowledge, there does not exist a closed-form solution for such points for general sets of vectors. However, programs for the relevant optimisation problem can be formulated and, for example, the methods represented in \cite{andersen1998minimizing} are reported to be able to handle $10^5$ parameters that would amount to roughly $N=16$ measurements. Remarkably, this high parameter count was reported with the computational standards of the year $2000$. However, as numerical search is not the main focus of this work, we leave open the question of how high numbers of parameters can be handled in our specific problem together with the question of how such a method would compare to the predominantly used numerical methods based on semi-definite programming in general. We can say, however, based on our numerical calculations, that the SOCP formulation of our condition performs similarly to SDP methods in the qubit case presented here. We suspect that such behaviour can be explained via an observation due to \cite{fawzi2019representing}. Namely, the second-order cone describes the semidefinite cone for a qubit. However, for qutrits this is not the case \cite{fawzi2019representing}, and a speed-up in computations could potentially be reached. This poses an interesting open question whether it is possible to obtain a characterization similar to ours in higher-dimensional systems, as well as for measurements with more outcomes.

\section{Acknowledgements} 

We thank Francisco Escudero Gutiérrez for valuable comments on the earlier version of the manuscript. RU is thankful for the financial support from the Swiss National Science Foundation (Ambizione PZ00P2-202179). DG was supported by SwissMAP scholarship and an NWO Vidi grant (Project No VI.Vidi.192.109).

\bibliography{references.bib} 

\section{Appendix}
\appendix

\section{\label{sec:proof} Proof of Theorems \ref{thm:necessary_condition}-\ref{thm:sufficient_condition}}

\begin{proposition} \label{prop:fourier}
Any joint POVM $\{J(\mu)\}_{\mu \in \Omega^N }$ for $N$ binary qubit observables, if it exists, can be written as
\begin{align} \label{prop:eq:main_form}
     J(\mu) &= \frac{1}{2^N} 
     \bigg( \bigg( 1 + \sum_{\substack{S \subseteq [N] \\ \abs{S} \geqslant 1}} z(S) \chi_S(\mu) \bigg) \id + \nonumber \\
     &+ \bigg( \sum_{\substack{S \subseteq [N] \\ \abs{S} \geqslant 1}} \bm{z}(S) \chi_S(\mu) \bigg) \cdot \vec{\sigma} \bigg),
\end{align}
where for every nonempty $S \subseteq [N]$ we define $\chi_S(\mu) \defeq \prod_{i \in S} \mu_i$.
\begin{proof} Any qubit POVM element $J(\mu)$ can be written as
\begin{equation}
    J(\mu) = \hat{z}(\mu) \id + \hat{\bm{z}}(\mu) \cdot \vec{\sigma}
\end{equation}
for some functions $\hat{z} : \Omega^N \rightarrow \mathbbm{R}$ and $\hat{\bm{z}} : \Omega^N \rightarrow \mathbbm{R}^3$.
It is a standard fact that any function can be expressed through its Fourier transformation \cite{de2008brief,o2014analysis} which in our case takes the form
\begin{equation}
    \hat{z}(\mu) = \frac{1}{2^N} \sum_{S \subseteq [N]} z(S) \chi_S(\mu),
\end{equation}
where $\chi_S(\mu) \defeq \prod_{i \in S} \mu_i$ for every nonempty $S \subseteq [N]$ and $\chi_\varnothing(\mu) \defeq 1$. The similar expression is true for vector-valued function $\hat{\bm{z}}(\mu)$. In order to get the form as in Eq.~(\ref{prop:eq:main_form}), we need to separate the coefficients that corresponds to the empty set. One condition of being POVM for the collection $\{J(\mu)\}_{\mu \in \Omega^N }$  necessary translates into 
\begin{equation}
    \sum_{\mu \in \Omega^N} J(\mu) = \id,
\end{equation}
implying
\begin{align}
    \sum_{\mu \in \Omega^N} \hat{z}(\mu) = 1 \text{ and } \sum_{\mu \in \Omega^N} \hat{\bm{z}}(\mu) = \bm{0}, 
\end{align}
which using the simple fact (e.g. see \cite{o2014analysis})
\begin{equation} \label{prop:sum_mu_fact}
    \frac{1}{2^N}\sum_{\mu \in \Omega^N} \chi_S(\mu) = 
    \begin{cases}
        1 &\text{ if $S = \varnothing$,}\\
        0 &\text{ if $S \neq \varnothing$},
    \end{cases}
\end{equation}
implies $z(\varnothing) = 1$ and $\bm{z}(\varnothing) = \bm{0}$.
\end{proof}
\end{proposition}

\begin{reusethm}{thm:necessary_condition}
If $N$ binary qubit observables are jointly measurable then
\begin{equation} 
    2^{N-1} \geq \min_{\cF_{N}} \sum_{\mathbf{\mu} \in \cD^{N}} \norm[\Bigg]{ \sum_{\substack{S \subseteq [N] \\ \text{odd }\abs{S}}} \bm{z}(S) \chi_S(\mu) },
\end{equation}
\end{reusethm}
\begin{proof}[Proof of Theorem \ref{thm:necessary_condition}]
According to Proposition~\ref{prop:fourier}, qubit POVM $J$, which represents a joint measurement, can be written as in Eq.~(\ref{prop:eq:main_form}). It has the property  $J(\mu) \geq 0$ for every $\mu \in \Omega^N$, which in the qubit case means that for every $\mu \in \Omega^N$:
\begin{align}
    1 + \sum_{\substack{S \subseteq [N] \\ \abs{S} \geqslant 1}} z(S) \chi_S(\mu)  \geq \norm[\Bigg]{ \sum_{\substack{S \subseteq [N] \\ \abs{S} \geqslant 1}} \bm{z}(S) \chi_S(\mu)  } .
\label{thm:eq:positivity}
\end{align}
Using the fact that complement of the set $\cD^{N}$ in $\Omega^N$ can be viewed as the set $\cD^{N}$ itself, where each coordinate of the vector $\mu$ is multiplied by $-1$, we can write
\begin{align} \label{thm:eq:compl}
    \sum_{\mu \in \Omega^N \backslash \cD^{N}} & \norm[\Bigg]{ \sum_{\substack{S \subseteq [N] \\ \abs{S} \geqslant 1}} \bm{z}(S) \chi_S(\mu)} = \sum_{-\mu \in \cD^{N}} \norm[\Bigg]{ \sum_{\substack{S \subseteq [N] \\ \abs{S} \geqslant 1}} \bm{z}(S) \chi_S(\mu)} \nonumber \\
    &= \sum_{\mu \in \cD^{N}} \norm[\Bigg]{ \sum_{\substack{S \subseteq [N] \\ \abs{S} \geqslant 1}} \bm{z}(S) \chi_S(-\mu)},
\end{align}
where $\chi_S(-\mu) = \chi_S(\mu)$ for even $\abs{S}$ and $\chi_S(-\mu) = - \chi_S(\mu)$ for odd $\abs{S}$. Summing over $\Omega^N$, it follows from Eqs.~(\ref{prop:sum_mu_fact}),~(\ref{thm:eq:positivity}),~(\ref{thm:eq:compl}) that 
\begin{align} \label{thm:eq:necessary}
    2^N &\geq \sum_{\mu \in \Omega^N} \norm[\Bigg]{ \sum_{\substack{S \subseteq [N] \\ \abs{S} \geqslant 1}} \bm{z}(S) \chi_S(\mu)} \nonumber \\
    &=
    \sum_{\mu \in \cD^{N}} \norm[\Bigg]{ \sum_{\substack{S \subseteq [N] \\ \text{odd }\abs{S}}} \bm{z}(S) \chi_S(\mu) + \sum_{\substack{S \subseteq [N] \\ \text{even }\abs{S}}} \bm{z}(S) \chi_S(\mu)  } \nonumber  \\
    &+ \mkern-10mu \sum_{\mu \in \Omega^N \backslash \cD^{N}} \norm[\Bigg]{ \sum_{\substack{S \subseteq [N] \\ \text{odd }\abs{S}}} \bm{z}(S) \chi_S(\mu)  + \sum_{\substack{S \subseteq [N] \\ \text{even }\abs{S}}} \bm{z}(S) \chi_S(\mu)  }  \nonumber \\
    &= 
     \sum_{\mu \in \cD^{N}} \norm[\Bigg]{ \sum_{\substack{S \subseteq [N] \\ \text{odd }\abs{S}}} \bm{z}(S) \chi_S(\mu)  + \sum_{\substack{S \subseteq [N] \\ \text{even }\abs{S}}} \bm{z}(S) \chi_S(\mu)  } \nonumber  \\
     &+ \sum_{\mu \in \cD^{N}} \norm[\Bigg]{\mkern5mu -\mkern-10mu \sum_{\substack{S \subseteq [N] \\ \text{odd }\abs{S}}} \bm{z}(S) \chi_S(\mu)  + \sum_{\substack{S \subseteq [N] \\ \text{even }\abs{S}}} \bm{z}(S) \chi_S(\mu)  } \nonumber \\
     &\geqslant 2 \sum_{\mu \in \cD^{N}} \norm[\Bigg]{ \sum_{\substack{S \subseteq [N] \\ \text{odd }\abs{S}}} \bm{z}(S) \chi_S(\mu)  },  
\end{align}
where we used triangle inequality in the last step. Note, that we could equivalently select the sum inside the absolute value to be defined over even $\abs{S}$. However, for the purpose of proving an interesting necessary condition, we need to choose odd $\abs{S}$ in triangle inequality, since all given qubit POVMs are specified by vectors $\bm{z}(1), \dotsc, \bm{z}(N)$ belonging to the odd $\abs{S}$ category. That gives us the desired condition:
\begin{align} \label{theorem1:final_eq}
    2^{N-1} &\geq \sum_{\mathbf{\mu} \in \cD^{N}} \norm[\Bigg]{ \sum_{\substack{S \subseteq [N] \\ \text{odd }\abs{S}}} \bm{z}(S) \chi_S(\mu)  } \nonumber \\
    &\geq \min_{\cF_{N}} \sum_{\mathbf{\mu} \in \cD^{N}} \norm[\Bigg]{ \sum_{\substack{S \subseteq [N] \\ \text{odd }\abs{S}}} \bm{z}(S) \chi_S(\mu)  }
\end{align}
\end{proof}

\begin{remark}
    The definition of $\cD^{N}$ might seem suspicious because of somewhat arbitrary condition $\mu_1 = +1$. In fact, there is a lot of freedom in the way one may define the set $\cD^{N}$. For example, one can define $\cD^{N}$ via fixing some arbitrary index $k \in [N]$ and setting $\mu_k = +1$. The proof and the claim of the theorem itself do not depend on this, because the corresponding norms in the sum of Eq.~(\ref{theorem1:final_eq}) are invariant to such choice. In general, one can define $\cD^{N}$ arbitrarily as long as the complement set $\Omega^N \backslash \cD^{N}$ is in one-to-one correspondence with $\cD^{N}$ under the involution $r :\Omega^N \rightarrow \Omega^N$, which acts as the reflection $r(\mu) = - \mu$ through the origin of the boolean cube, such that $r(\cD^{N}) = \Omega^N \backslash \cD^{N}$. However, our specific choice of the set $\cD^{N}$ allows in the following to prove Theorem~\ref{thm:sufficient_condition} more easily (in particular, see Proposition~\ref{thm2:prop1}).
\end{remark}

\begin{reusethm}{thm:sufficient_condition}
If for $N$ binary unbiased qubit observables, specified by vectors $\bm{z}^\prime(1),\dotsc,\bm{z}^\prime(N)$, there exist vectors $\bm{z}^\prime(S) \in \cF_{N}$ for every corresponding odd $\abs{S} > 1$, such that
\begin{equation} \label{eq:sufficient_condition_2}
    2^{N-1} \geq \sum_{\mathbf{\mu} \in \cD^{N}} \norm[\Bigg]{ \sum_{\substack{S \subseteq [N] \\ \text{odd }\abs{S}}} \bm{z}^\prime(S) \chi_S(\mu) },
\end{equation}
then these $N$ qubit observables are jointly measurable.
\end{reusethm}
\begin{proof}[Proof of Theorem \ref{thm:sufficient_condition}]
We are given a set of vectors $\{\bm{z}^\prime(S)\}_{\abs{S} \text{ odd}}$ with the property~(\ref{eq:sufficient_condition_2}). We will construct an ansatz for a joint measurement $\{J(\mu)\}_{\mu \in \Omega^N }$ by choosing specific vectors $\{\bm{z}(S)\}_{S \subseteq [N]}$ and scalars $\{z(S)\}_{S \subseteq [N]}$ as in Eq.~(\ref{prop:eq:main_form}). Let for every index set $S \subseteq [N]$ we define a $\mathbbm{R}^4$-vector $\left(z(S), \bm{z}(S) \right)$ as
\begin{align} \label{thm2:def_vec_scal}
    \left(z(S), \bm{z}(S) \right) &\defeq \begin{cases}
        \left(0,\bm{z}^\prime(S) \right) &\text{ if $\abs{S}$ is odd,}\\
        \left(z^\prime(S),\bm{0}\right) &\text{ if $\abs{S}$ is even,}
    \end{cases}
\end{align}
where $z^\prime(\varnothing) \defeq 1$ and $\{z^\prime(S)\}_{\text{even }\abs{S}\geqslant 2}$ is the solution of a system of linear equations:
\begin{align} \label{eq:lin_sys_orig}
     1 + \sum_{\substack{S \subseteq [N] \\ \mathclap{ \text{even }\abs{S} \geqslant 2}}} z^\prime(S) \chi_S(\mu) = \norm[\Bigg]{ \sum_{\substack{S \subseteq [N] \\  \text{odd }\abs{S}}} \bm{z}^\prime(S) \chi_S(\mu) },
\end{align}
where $\mu \in \cD^{N} \backslash \{\mu_+\}$. The system above is always solvable. This is the content of the Lemma~\ref{lemma:lin_sys} and Propositions~\ref{thm2:prop1}-\ref{thm2:prop2}, which we prove right after the current theorem. Assume for now that Lemma~\ref{lemma:lin_sys} and Propositions~\ref{thm2:prop1}-\ref{thm2:prop2} are true. Note that the definition of $z(S)$ for odd $\abs{S}$ reflects the assumption that our qubit observables are unbiased. 

We need to show that the ansatz for $J(\mu)$ defines a valid POVM, i.e. we need to verify that $\sum_{\mu \in \Omega^N} J(\mu) = \id$ and that for every $\mu \in \Omega^N$ the operator $J(\mu) \geq 0$.
The way we defined the vectors and scalars in Eq.~(\ref{thm2:def_vec_scal}) allows us to write
\begin{align} \label{thm2:J_ansatz}
    &J(\mu) = \frac{1}{2^N}
     \bigg( \bigg( 1 + \sum_{\substack{S \subseteq [N] \\ \abs{S} \geqslant 1}} z(S) \chi_S(\mu) \bigg) \id  \nonumber \\
     &+ \bigg( \sum_{\substack{S \subseteq [N] \\ \abs{S} \geqslant 1}} \bm{z}(S) \chi_S(\mu) \bigg) \cdot\vec{\sigma} \bigg) \nonumber \\
     &= \frac{1}{2^N}
     \bigg( \bigg( 1 + \sum_{\substack{S \subseteq [N] \\ \mathclap{ \text{even }\abs{S} \geqslant 2}}} z^\prime(S) \chi_S(\mu)  \bigg)\id \nonumber \\
     &+ \bigg( \sum_{\substack{S \subseteq [N] \\  \text{odd }\abs{S}}} \bm{z}^\prime(S) \chi_S(\mu)  \bigg) \cdot\vec{\sigma} \bigg).
\end{align}
Note that from the first line above in Eq.~(\ref{thm2:J_ansatz}) it is clear that $\sum_{\mu \in \Omega^N} J(\mu) = \id$:
\begin{align} \label{lem1:sum_to_iden}
    &\sum_{\mu \in \Omega^N} J(\mu) = \frac{1}{2^N} \bigg( \bigg( \sum_{\mu \in \Omega^N} 1 + \sum_{\substack{S \subseteq [N] \\ \abs{S} \geqslant 1}} z(S) \sum_{\mu \in \Omega^N} \chi_S(\mu) \bigg) \id \nonumber \\
    &+\bigg( \sum_{\substack{S \subseteq [N] \\ \abs{S} \geqslant 1}} \bm{z}(S) \sum_{\mu \in \Omega^N} \chi_S(\mu) \bigg) \cdot \vec{\sigma} \bigg) \nonumber \\
    &= \frac{1}{2^N} \bigg( \bigg( 2^N + \sum_{\substack{S \subseteq [N] \\ \abs{S} \geqslant 1}} z(S) \cdot 0  \bigg) \id + \bigg( \sum_{\substack{S \subseteq [N] \\ \abs{S} \geqslant 1}} \bm{z}(S) \cdot 0  \bigg) \cdot \vec{\sigma} \bigg) \nonumber \\
    &= \id,  
\end{align}
where we used Eq.~(\ref{prop:sum_mu_fact}).
So we need to show that for every $\mu \in \Omega^N$ the operator $J(\mu)$ is positive semidefinite, which using the second line in Eq.~(\ref{thm2:J_ansatz}) is equivalent to the statement that for every $\mu \in \Omega^N$:
\begin{align} \label{thm2:pos_cond}
     1 + \sum_{\substack{S \subseteq [N] \\ \mathclap{ \text{even }\abs{S} \geqslant 2}}} z^\prime(S) \chi_S(\mu) \geqslant \norm[\Bigg]{ \sum_{\substack{S \subseteq [N] \\  \text{odd }\abs{S}}} \bm{z}^\prime(S) \chi_S(\mu) }.
\end{align}
For all vectors $\mu \in \cD^{N} \backslash \{\mu_+\}$ the condition above holds automatically by construction of the linear system~(\ref{eq:lin_sys_orig}). Let us verify it for $\mu = \mu_+$. For that the sum of Equations~(\ref{eq:lin_sys_orig}) can be written as
\begin{align}
     &\sum_{\mu \in \cD^{N} \backslash \{\mu_+\}} \norm[\Bigg]{ \sum_{\substack{S \subseteq [N] \\  \text{odd }\abs{S}}} \bm{z}^\prime(S) \chi_S(\mu) } = \nonumber \\
     &=  \sum_{\mu \in \cD^{N} \backslash \{\mu_+\}} \bigg( 1 + \sum_{\substack{S \subseteq [N] \\ \mathclap{ \text{even }\abs{S} \geqslant 2}}} z^\prime(S) \chi_S(\mu) \bigg) \nonumber \\
     &= \abs{\cD^{N} \backslash \{\mu_+\}} + \sum_{\substack{S \subseteq [N] \\ \mathclap{ \text{even }\abs{S} \geqslant 2}}} z^\prime(S) \sum_{\mu \in \cD^{N} \backslash \{\mu_+\}}\chi_S(\mu)  \nonumber \\
     &=  2^{N-1} - 1 - \sum_{\substack{S \subseteq [N] \\ \mathclap{ \text{even }\abs{S} \geqslant 2}}} z^\prime(S)
\end{align}
where in the third equality we used Proposition~\ref{thm2:prop1} and the fact $\abs{\cD^{N} \backslash \{\mu_+\}} = 2^{N-1} - 1$ (see Remark~\ref{remark_3}). Therefore for $\mu = \mu_+$ we have
\begin{align} 
     1 + &\sum_{\substack{S \subseteq [N] \\ \mathclap{ \text{even }\abs{S} \geqslant 2}}} z^\prime(S) \chi_S(\mu_+) = 1 + \sum_{\substack{S \subseteq [N] \\ \mathclap{ \text{even }\abs{S} \geqslant 2}}} z^\prime(S) \nonumber \\
     &= 2^{N-1} - \mkern-20mu \sum_{\mathbf{\mu} \in \cD^{N}\backslash \{\mu_+\}} \norm[\Bigg]{ \sum_{\substack{S \subseteq [N] \\ \text{odd }\abs{S}}} \bm{z}(S) \chi_S(\mu)}  \nonumber \\
     &\geq 
     \norm[\Bigg]{\sum_{\substack{S \subseteq [N] \\ \text{odd }\abs{S}}} \bm{z}(S) \chi_S(\mu_+)},
\end{align}
where in the last inequality we used the assumption (\ref{eq:sufficient_condition_2}). So we see that Eq.~(\ref{thm2:pos_cond}) holds for every $\mu \in \cD^{N}$. Finally, it is easy to see that it also holds for every $\mu \in \Omega^N \backslash \cD^{N}$, because $-\mu \in \cD^{N}$ is equivalent to $\mu \in \Omega^N \backslash \cD^{N}$. So we can rewrite Eq.~(\ref{thm2:pos_cond}) for every $-\mu \in \cD^{N}$ as
\begin{align*} 
     1 + \sum_{\substack{S \subseteq [N] \\ \mathclap{ \text{even }\abs{S} \geqslant 2}}} z^\prime(S) \chi_S(-\mu) \geq \norm[\Bigg]{ \sum_{\substack{S \subseteq [N] \\  \text{odd }\abs{S}}} \bm{z}^\prime(S) \chi_S(-\mu) }
\end{align*}
for every $\mu \in \Omega^N \backslash \cD^{N}$. Using $\chi_S(-\mu) = \chi_S(\mu)$ for even $\abs{S}$ and $\chi_S(-\mu) = - \chi_S(\mu)$ for odd $\abs{S}$, it is equivalent to
\begin{align} 
     1 + \sum_{\substack{S \subseteq [N] \\ \mathclap{ \text{even }\abs{S} \geqslant 2}}} z^\prime(S) \chi_S(\mu) \geq \norm[\Bigg]{ \sum_{\substack{S \subseteq [N] \\  \text{odd }\abs{S}}} \bm{z}^\prime(S) \chi_S(\mu) }
\end{align}
for every $\mu \in \Omega^N \backslash \cD^{N}$.
\end{proof}

\begin{remark}
We actually proved more in Theorem~\ref{thm:sufficient_condition} than just an existence of the joint measurement. We explicitly constructed the specific joint POVM $\{J(\mu)\}_{\mu \in \Omega^N }$ in the proof via Eq.~(\ref{thm2:def_vec_scal}), linear system (\ref{eq:lin_sys_orig}) and Eq.~(\ref{thm2:J_ansatz}).
\end{remark}

\begin{remark} \label{remark_3} For the Lemma~\ref{lemma:lin_sys} and Propositions~\ref{thm2:prop1}-\ref{thm2:prop2} it is convenient to rewrite the linear system (\ref{eq:lin_sys_orig}) in the form 
\begin{equation} \label{eq:lin_sys}
    A \mathbf{x} = \mathbf{b},
\end{equation}
where $\mathbf{x}$ is the vector with components $z^\prime(S)$ (except for $S = \varnothing$) written in lexicographic order, i.e. the vector looks like $ \mathbf{x} = \left(z^\prime(12),\dotsc,z^\prime(N-1,N),z^\prime(1234),\dotsc \right)$, where $z^\prime(12)$ should be read as $z^\prime(\{1,2\})$ and so on. We define $\mathcal{X}$ to be its ordered list of labels $S$ of the components. For example, when $N=3$ the label set $\mathcal{X} = \{ \{1,2\},\{1,3\},\{2,3\}\}$. The vector $\mathbf{x}$ has all possible even combinations of distinct indices, and has therefore the length
\begin{equation*}
    d(N) \defeq \sum_{k=1}^{\floor{\frac{N}{2}}} \binom{N}{2k} =  2^{N-1}-1.
\end{equation*}
The vector $\mathbf{b}$ has components $b(\mu)$ defined for every $\mu \in \cD^{N} \backslash \{\mu_+\}$ as 
\begin{equation}
    b(\mu) \defeq \norm[\Bigg]{\sum_{\substack{S \subseteq [N] \\\text{odd } \abs{S} }} \bm{z}^\prime(S) \chi_S(\mu) } - 1,
\end{equation}
which are arranged in lexicographic order with respect to $-1$ in the vector $\mu$. For example, when $N=4$ then
\begin{align*}
    &\mathbf{b} = \left( b(+,-,+,+),b(+,+,-,+),b(+,+,+,-), \right. \\ & \left.  b(+,-,-,+),b(+,-,+,-),b(+,+,-,-),b(+,-,-,-) \right) ,
\end{align*}
which, using a different label convention where the arguments enumerate positions of $-1$ in the vector $\mu$, we can rewrite in a more compact form as
\begin{equation}
    \mathbf{b} = \left(b(2),b(3),b(4),b(23),b(24),b(34),b(234)\right),
\end{equation} 
where $b(12)$ should be read as $b(\{1,2\})$ and so on. Also let $\mathcal{B}$ be the ordered list of labels of the components of $\mathbf{b}$. Note that $\mathcal{B}$ has the cardinality $\abs{\cD^{N} \backslash \{\mu_+\}}$, which clearly equals $d(N)$. That means that the matrix $A$ is a square $d(N) \times d(N)$ matrix. The matrix $A$ has only entries from $\{-1,+1\}$. For example, when $N=4$ label sets are $\mathcal{X} = \{ \{1,2\},\{1,3\},\{1,4\},\{2,3\},\{2,4\},\{3,4\},\{1,2,3,4\}\}$ and $\mathcal{B} = \{\{2\},\{3\},\{4\},\{2,3\},\{2,4\},\{3,4\},\{2,3,4\}\}$, and $A$ looks like
\begin{equation*}
    A = 
    \kbordermatrix{& 12 & 13 & 14 & 23 & 24 & 34 & \mathclap{1234} \\
    2 & - & + & + & - & - & + & - \\
    3 & + & - & + & - & + & - & - \\
    4 & + & + & - & + & - & - & - \\
    23 & - & - & + & + & - & - & + \\
    24 & - & + & - & - & + & - & + \\
    34 & + & - & - & - & - & + & + \\
    234 & - & - & - & + & + & + & -
    }
\end{equation*}
In general, the matrix entry $A_{k i}$ can be calculated as $\chi_S(\mu)= (-1)^{\abs{\mathcal{B}_k \cap \mathcal{X}_i}}$, where $\mathcal{B}_k \cap \mathcal{X}_i$ denotes the intersection of the corresponding two labels from the labels sets, e.g. $\mathcal{B}_6 \cap \mathcal{X}_5 = \{ 4 \}$ for $N=4$. This is because the set $\mathcal{B}_k \cap \mathcal{X}_i$ enumerates exactly the components of vector $\mu$, which are $-1$ and appear in the product $\chi_{S}(\mu)$ near corresponding scalar $z^\prime(S)$ in Eq.~(\ref{eq:lin_sys_orig}) (in other words, the label set is $S = \mathcal{X}_i$ and $\mu$ corresponds to $\mathcal{B}_k$).
\end{remark} 

\begin{lemma} \label{lemma:lin_sys}
    Linear system of equations (\ref{eq:lin_sys_orig}) always has unique solution.
\end{lemma}
\begin{proof}
Consider the rewriting of the linear system as in Eq.~(\ref{eq:lin_sys}). For the proof we construct an ansatz for the inverse matrix $A^{-1}$ explicitly, and show that it is indeed the inverse. Define $A^{-1}$ as follows:
\begin{equation}
    A^{-1}_{ij} \defeq \frac{1}{2^{N-2}} \cdot
    \begin{cases}
        0 &\text{if $A_{ji}=+1$} \\
        -1 &\text{if $A_{ji}=-1$} 
    \end{cases}
\end{equation}
Now let us calculate the entry $(A^{-1}A)_{ij}$:
\begin{align}
    &(A^{-1}A)_{ij} = \sum_{k=1}^{d(N)} A^{-1}_{ik}A_{kj} = \sum_{k=1}^{d(N)} \frac{A_{ki}-1}{2^{N-1}}A_{kj} \nonumber \\
    &= \frac{1}{2^{N-1}} \bigg(\sum_{k=1}^{d(N)} A_{ki}A_{kj} - \sum_{k=1}^{d(N)} A_{kj} \bigg) \nonumber \\
    &= \frac{1}{2^{N-1}} \bigg( \sum_{k=1}^{d(N)} (-1)^{\abs{\mathcal{B}_k \cap \mathcal{X}_i}+\abs{\mathcal{B}_k \cap \mathcal{X}_j}} - \sum_{k=1}^{d(N)} (-1)^{\abs{\mathcal{B}_k \cap \mathcal{X}_j}} \bigg) 
\end{align}
Now we use Proposition~\ref{thm2:prop1} to calculate the second sum and Proposition~\ref{thm2:prop2} to calculate the first sum: 
\begin{align}
    (A^{-1}A)_{ij} = 1 - \frac{\mathscr{D}_H(A_{\cdot i},A_{\cdot j})}{2^{N-2}} = \delta_{ij}.
\end{align}
Therefore, the solution $\mathbf{x}$ exists, is unique and equals $\mathbf{x} = A^{-1}\mathbf{b}$.
\end{proof}

\begin{proposition} \label{thm2:prop1}
For every column $j$ of the matrix $A$ the sum of its entries $\sum_i A_{ij}$ equals
\begin{equation}
     \sum_{\mu \in \cD^{N} \backslash \{\mu_+\}}\chi_S(\mu) = \sum_{i=1}^{d(N)} (-1)^{\abs{\mathcal{B}_i \cap \mathcal{X}_j}} = -1,
\end{equation}
where $S=\mathcal{X}_j$ and label sets $\mathcal{B}_i \in \mathcal{B}$ correspond to vectors $\mu \in \cD^{N} \backslash \{\mu_+\}$ as in Remark~(\ref{remark_3}).
\end{proposition}
\begin{proof}
For each $k \in [N-1]$ we define a subset $\mathcal{B}^{k} \subset \mathcal{B}$ to be the set of labels of size $k$. The cardinality of this subset is $\binom{N-1}{k}$. We will denote the labels from this subset by $\mathcal{B}^{k}_{i}$ for some $i \in \left[\binom{N-1}{k}\right]$. Fix some column $j$ of the matrix $A$, which corresponds to a label set $\mathcal{X}_{j}$. The cardinality of this label is even and we will denote it by $2l \defeq \abs{\mathcal{X}_{j}}$. Now we need to differentiate between two cases.
\begin{enumerate}
\item The index $1 \not\in \mathcal{X}_j$. We can rewrite our sum  as
\begin{align}
    \sum_{i=1}^{d(N)} (-1)^{\abs{\mathcal{B}_i \cap \mathcal{X}_j}} = \sum_{k=1}^{N-1} \sum_{i=1}^{\binom{N-1}{k}} (-1)^{\abs{\mathcal{B}^{k}_i \cap \mathcal{X}_j}}.
\end{align}
But for fixed $k$ we can write
\begin{equation}
    \sum_{i=1}^{\binom{N-1}{k}} (-1)^{\abs{\mathcal{B}^{k}_i \cap \mathcal{X}_j}} 
    = \sum_{m=0}^{2l} \binom{2l}{m} \binom{N-1-2l}{k-m} (-1)^m,
\end{equation}
where we define binomial coefficients to be zero $\binom{x}{y} = 0$ if $y>x$ or $y<0$. This is justified as follows: $m$ is the number of elements in the intersection $\mathcal{B}^{k}_i \cap \mathcal{X}_j$, the first binomial coefficient counts the number of choices of these $m$ common elements in the set $\mathcal{X}_j$, and the second coefficient counts all the different sets $\mathcal{B}^{k}_i$ with these and only these $m$ common elements. Therefore
\begin{align}
    \sum_{i=1}^{d(N)} &(-1)^{\abs{\mathcal{B}_i \cap \mathcal{X}_j}}
    =  \sum_{k=1}^{N-1} \sum_{m=0}^{2l}  \binom{2l}{m} \binom{N-1-2l}{k-m} (-1)^m 
    \nonumber \\
    &= \sum_{m=0}^{2l} \binom{2l}{m} (-1)^m \sum_{k=1}^{N-1}  \binom{N-1-2l}{k-m} \nonumber \\
    &= \sum_{k=1}^{N-1} \binom{N-1-2l}{k} + \\
    &+ \sum_{m=1}^{2l} \binom{2l}{m} (-1)^m \sum_{k=1}^{N-1}  \binom{N-1-2l}{k-m} \nonumber \\
    &= \left(2^{N-1-2l} -1 \right) + \sum_{m=1}^{2l} \binom{2l}{m} (-1)^m \cdot 2^{N-1-2l} \nonumber \\
    &= 2^{N-2l-1} -1 + \bigg( \bigg( \sum_{m=0}^{2l} \binom{2l}{m}(-1)^m \bigg) - 1 \bigg) \cdot 2^{N-1-2l} \nonumber \\
    &= 2^{N-2l-1} - 1 + \left( 0 - 1 \right) \cdot 2^{N-1-2l} = -1. \nonumber
\end{align}
Note, that the final result does not depend on $j$.
\item The index $1 \in \mathcal{X}_j$. We employ a similar argument as in the first case. The only difference is that now we have $\binom{2l-1}{m}$ choices of $m$ elements in the label $\mathcal{X}_j$ for the intersection $\mathcal{B}^{k}_i \cap \mathcal{X}_j$, because the label $\mathcal{B}^{k}_i$ never contains number $1$. The rest $k-m$ numbers comprising the label $\mathcal{B}^{k}_i$ are selected from $[N] \backslash \mathcal{X}_j$. Therefore
\begin{align}
    \sum_{i=1}^{d(N)} &(-1)^{\abs{\mathcal{B}_i \cap \mathcal{X}_j}} = \sum_{k=1}^{N-1} \sum_{i=1}^{\binom{N-1}{k}} (-1)^{\abs{\mathcal{B}^{k}_i \cap \mathcal{X}_j}} \nonumber \\ 
    &= \sum_{k=1}^{N-1} \sum_{m=0}^{2l-1} \binom{2l-1}{m} \binom{N-2l}{k-m} (-1)^m
\end{align}
Now we can simplify this expression as in the first case:
\begin{align}
    \sum_{i=1}^{d(N)} &(-1)^{\abs{\mathcal{B}_i \cap \mathcal{X}_j}}
    =  \sum_{k=1}^{N-1} \sum_{m=0}^{2l-1}  \binom{2l-1}{m} \binom{N-2l}{k-m} (-1)^m 
    \nonumber \\
    &= \sum_{m=0}^{2l-1} \binom{2l-1}{m} (-1)^m \sum_{k=1}^{N-1}  \binom{N-2l}{k-m} \\
    &= \sum_{k=1}^{N-1} \binom{N-2l}{k} + \sum_{m=1}^{2l-1} \binom{2l-1}{m} (-1)^m \sum_{k=1}^{N-1}  \binom{N-2l}{k-m} \nonumber \\
    &= \left(2^{N-2l} -1 \right) + \sum_{m=1}^{2l-1} \binom{2l-1}{m} (-1)^m \cdot 2^{N-2l} \nonumber \\
    &= 2^{N-2l} - 1 + \bigg( \bigg( \sum_{m=0}^{2l-1} \binom{2l-1}{m}(-1)^m \bigg) - 1 \bigg) \cdot 2^{N-2l} \nonumber \\
    &= 2^{N-2l} - 1 + \left( 0 - 1 \right) \cdot 2^{N-2l} = -1. \nonumber
\end{align}
Again, the final result does not depend on $j$.
\end{enumerate}
\end{proof}

\begin{proposition} \label{thm2:prop2}
Let $\mathscr{D}_H(A_{\cdot i},A_{\cdot j})$ be the Hamming distance between columns $i$ and $j$ in the matrix $A$, i.e. the number of corresponding coordinates of two column vectors, which are not equal. Then for every $i$ and $j$ it is true that
\begin{equation}
    \mathscr{D}_H(A_{\cdot i},A_{\cdot j}) = \begin{cases}
        0 &\text{ if $i=j$,} \\
        2^{N-2} &\text{ if $i\neq j$,}
    \end{cases} 
\end{equation}
Moreover, this is equivalent to 
\begin{equation}
\sum_{k=1}^{d(N)} (-1)^{\abs{\mathcal{B}_k \cap \mathcal{X}_i}+\abs{\mathcal{B}_k \cap \mathcal{X}_j}} = \begin{cases}
        2^{N-1}-1 &\text{ if $i=j$,} \\
        -1 &\text{ if $i\neq j$.}
    \end{cases} 
\end{equation}
\end{proposition}
\begin{proof}
The case $i=j$ is trivial. When $i \neq j$, fix some column vectors $A_{\cdot i}, A_{\cdot j}$, specified by some labels $\mathcal{X}_i,\mathcal{X}_j$ respectively. Let us look at $k$-th components of these two column vectors, which correspond to the label $\mathcal{B}_k$. Observe that, $k$-th coordinates of two column vectors $A_{\cdot i}$ and $A_{\cdot j}$ are the same if and only if $\abs{\mathcal{B}_k \cap \mathcal{X}_i}$ and $\abs{\mathcal{B}_k \cap \mathcal{X}_j}$ has the same parity, and vice versa. That means we can study the sum $\sum_{k=1}^{d(N)} (-1)^{\abs{\mathcal{B}_k \cap \mathcal{X}_i}+\abs{\mathcal{B}_k \cap \mathcal{X}_j}}$ instead of $\mathscr{D}_H(A_{\cdot i},A_{\cdot j})$:
\begin{align} \label{thm2:prop2_sum_hamming}
    & \sum_{k=1}^{d(N)} (-1)^{\abs{\mathcal{B}_k \cap \mathcal{X}_i} + \abs{\mathcal{B}_k \cap \mathcal{X}_j}} = \\
    &= -\mathscr{D}_H(A_{\cdot i},A_{\cdot j}) + \left( d(N) - \mathscr{D}_H(A_{\cdot i},A_{\cdot j}) \right)
    \nonumber \\ 
    &= 2^{N-1}-1 - 2 \mathscr{D}_H(A_{\cdot i},A_{\cdot j}). \nonumber
\end{align}
Note that $\mathcal{B}_k \cap \mathcal{X}_i = \left(\mathcal{B}_k \cap \mathcal{X}_i \cap \mathcal{X}_j \right) \cup \left(\mathcal{B}_k \cap \left(\mathcal{X}_i \backslash \mathcal{X}_j\right) \right)$ and the similar identity is true for $\mathcal{X}_j$. Therefore
\begin{align} \label{thm2:prop2:sum_hamming_2}
    &\sum_{k=1}^{d(N)} (-1)^{\abs{\mathcal{B}_k \cap \mathcal{X}_i}+\abs{\mathcal{B}_k \cap \mathcal{X}_j}} = \nonumber \\
    &= \sum_{k=1}^{d(N)} (-1)^{2 \cdot \abs{\mathcal{B}_k \cap \mathcal{X}_i \cap \mathcal{X}_j}+\abs{\mathcal{B}_k \cap \left(\mathcal{X}_i \backslash \mathcal{X}_j\right)} + \abs{\mathcal{B}_k \cap \left(\mathcal{X}_j \backslash \mathcal{X}_i\right)}} \nonumber \\
    &=  \sum_{k=1}^{d(N)} (-1)^{\abs{\mathcal{B}_k \cap \left(\mathcal{X}_i \bigtriangleup \mathcal{X}_j\right)}},
\end{align}
where $\mathcal{X}_i \bigtriangleup \mathcal{X}_j$ is the symmetric difference of the two sets $\mathcal{X}_i$ and $\mathcal{X}_j$. But since both these sets have an even number of elements, it follows that the set $\mathcal{X}_i \bigtriangleup \mathcal{X}_j$ also has even number of elements not equal to zero, i.e. $\mathcal{X}_i \bigtriangleup \mathcal{X}_j \in \mathcal{X}$, and this label corresponds to some other column, different from $\mathcal{X}_i$ and $\mathcal{X}_i$ in the matrix $A$. Therefore we can apply the Proposition~\ref{thm2:prop1} to conclude
\begin{align} \label{thm2:prop2:sum_hamming_3}
\sum_{k=1}^{d(N)} (-1)^{\abs{\mathcal{B}_k \cap \mathcal{X}_i}+\abs{\mathcal{B}_k \cap \mathcal{X}_j}} = \sum_{k=1}^{d(N)} (-1)^{\abs{\mathcal{B}_k \cap \left(\mathcal{X}_i \bigtriangleup \mathcal{X}_j\right)}} = -1.
\end{align}
Finally, from Eq.~(\ref{thm2:prop2_sum_hamming}) we get $\mathscr{D}_H(A_{\cdot i},A_{\cdot j}) = 2^{N-2}$ when $i \neq j$.
\end{proof}

\end{document}